\documentclass[%
 reprint,
 amsmath,amssymb,
 aps,
]{revtex4-2}

\usepackage{graphicx}
\usepackage{float}
\usepackage{dcolumn}
\usepackage{bm}
\usepackage{upgreek}

\usepackage{cleveref}   
\usepackage{color}
\usepackage{subcaption}
\usepackage{caption}
\begin{document}


\title{A Matter of Shape: Contact Area Optimization in Soft Lubrication}

\author{Joaquin Garcia-Suarez}
            \altaffiliation{Institute of Civil Engineering,
            \'{E}cole Polytechnique F\'{e}d\'{e}rale de Lausanne (EPFL), CH 1015 Lausanne, 
            Switzerland}


\begin{abstract}
We study the fluid-mediated approach of a deformable axisymmetric object towards a rigid substrate, focusing on how its shape influences contact formation. 
For low approach velocities and large Stokes numbers, 
we show that 
sharper profiles (e.g., conical) maximize contact at the center and avoid fluid entrapment, while blunter ones form central dimples that trap bubbles. 
We also find that the resulting pressure distributions in the presence of thin viscous films can be predicted remarkably well by classical (dry) contact mechanics. 
These findings reveal shape as a design parameter for contact optimization in soft matter, adhesion, and elastohydrodynamics. 
Finally, we also theorize the possibility of a mechanical equivalence between shape and approach velocity. 
\end{abstract}
\maketitle

Understanding how deformable objects establish contact with solid surfaces is crucial in a wide range of applications, from biomedical adhesives and surgical tools~\cite{chansoria:2024} to soft robotics~\cite{Wang:2022, Qin:2024} and engineered gripping systems~\cite{Wang:2022, yue:2024}. In many of these contexts, maximizing the area of solid-solid contact is desirable for adhesion, force transmission, or sensing. However, when a soft object approaches a rigid surface through an interstitial fluid \cite{Mahadevan:2004,Rallabandi:2024} such as air or water, lubrication pressures may develop, often preventing full contact and leading to the entrapment of a fluid pocket, as it is well-known to happen for droplets \cite{Bouwhuis:2012,riboux:2014,kolinski:2014,sprittles:2024}. 

\begin{figure}[t]
\centering
\includegraphics[width=\linewidth]{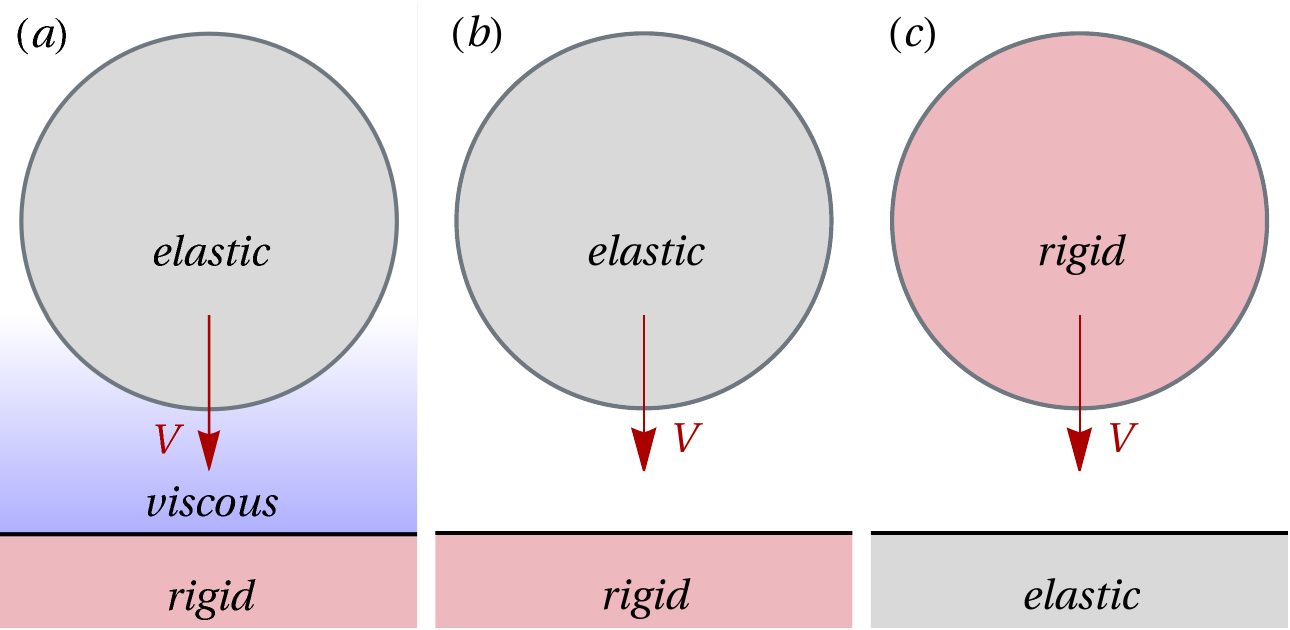}
\caption{
Schematic of the axisymmetric approach problem. 
(a) A deformable elastic indenter approaches a rigid substrate through a viscous fluid film. moving at initial velocity $V$. 
(b) Equivalent dry contact setup: a rigid indenter compressing an elastic half-space, equivalent to (a) in the limit $\mathrm{St} \to \infty$. 
(c) Role-reversed system: elastic indenter and rigid substrate without the fluid layer, it develops similar interface tractions as (b) but the deformation is not flattening of the indenter but indetation of the half-space. 
}
\label{fig:schematic}
\end{figure}

This fluid-mediated cushioning effect has been reported not only during drop impact, more recently also in the approach of compliant solids~\cite{Zheng:2021}, where viscous air pressures deform the soft tip and lead to air bubble entrapment. 
Theoretical and numerical work has shown that, under quasi-static conditions, the resulting elastohydrodynamic lubrication pressure distribution may resemble that predicted by classical dry contact mechanics~\cite{bertin:2024, Bilotto:2024}. 
One may wonder if these observations, reported for parabolic indenters~\cite{Johnson:1985, bertin:2024,Bilotto:2024}, would hold for other shapes.

\begin{figure}[H]
    \centering
    \includegraphics[width=0.95\linewidth]{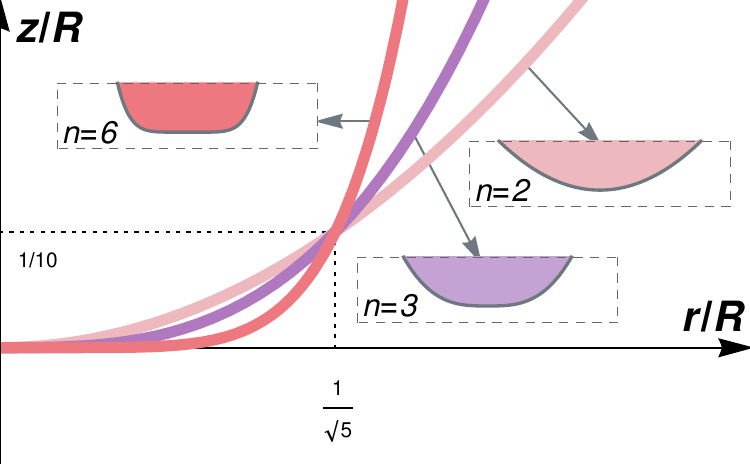}
    \caption{
    Three indenter profiles match at $r = R / \sqrt{5}$ and $z = h(r) = R/10$. Note: the three insets are at the same scale.
    }
    \label{fig:indenters}
\end{figure}

Here, we investigate how the shape of the approaching indenter affects the onset of contact and the final contact area in a fluid-mediated impact. 
%
%
%
%
Consider a soft, axisymmetric indenter, with a profiles of the form $f(r) = r^n /2k$ \cite{Popov:2019} ($k$ is a geometrical parameter with suitable units), moving towards a rigid substrate through a viscous fluid layer at an uniform initial velocity $V$, \Cref{fig:schematic}(a). 
The air film pressure obeys the Reynolds equation, while the solid deforms according to elastic integral kernels~\cite{David:1986}, i.e., 
we assume quasi-static deformation. Then, the coupled elastohydrodynamic problem is solved numerically using a modified version of the method developed in Refs.~\cite{liu:2022,bertin:2024}, adapted to handle profiles $\sim r^n$ for $n$ (see Supplementary Materials for full details on the method). 
The object is initialized at uniform velocity away from the substrate and it decelerates due to lubrication pressures as it tries to close the gap. 
The latter never fully closes in the simulations (such event would require other physics not included in the model, e.g., gas rarefaction), and eventually the approach stops, the velocity of the center of mass changes sign and the rebound phase would begin. 
The simulation is stopped at the timestep when the velocity changes sign. 
The governing equations are solved in dimensionless form, using as characteristic values associated to the fluid-less (``dry'') contact mechanics impact problem \cite{Popov:2019}:
\begin{align*}
    \mathcal{H} 
       & = 
        \left(
            \left( 
                {\kappa(n) \over 2 k} 
            \right)^{1/n}
            { (2n+1) (n+1)
            \over
            4 n^2}
            \frac{m V_0^2}{E^*}
        \right)^{\frac{n}{2n+1}} \, , \\
        \mathcal{L} 
        &= 
        \left(
            k \mathcal{H}
        \right)^{1/n} \, , 
        \quad
        \mathcal{P} 
        = 
        {2 E^* \over \pi}
        {\mathcal{H} \over \mathcal{L}} \, , 
        \quad
        \tau
        = 
        {m V_0 
            \over 
        E^* \mathcal{L} \mathcal{H}} \, ,
        \hfill 
\end{align*}
where $\mathcal{H}$ is the vertical length scale, $\mathcal{L}$ is the radial one, $\mathcal{P}$ is the characteristic pressure and $\tau$ thecharacteristicc time. 
Note that all of them are shape-dependent, i.e., functions of $n$, so the quantitative comparison among parameters of interest can not be done before transforming back to physical scales. 
$\kappa(n)$ is an order-one parameter depending solely on the profile exponent, termed ``stretch factor'' \cite{Popov:2019}. 
The Stokes number of the approach problem is thus expressed as
$\mathrm{St} = {\tau \mathcal{P} \mathcal{H}^2 / 12 \eta \mathcal{L}^2}$. 
$\mathrm{St} \gg 1$ also means that we also approach the dry limit, in which the effect of the fluid could be neglected and it would not delay contact between the solid and the substrate (\Cref{fig:indenters}(b)). 
By virtue of Johnson's analogy \cite{Johnson:1985}, the contact tractions that develop at the interface between the object and the substrate are the same for both the system in \Cref{fig:indenters}(b) and in \Cref{fig:indenters}(c); 
in the latter case they lead to denting on the half-space, while in the former to flattening of the leading edge. 
All simulations are performed at $\mathrm{St} = 1000$, ensuring global inertia controls the overall timescale while locally around the leading edge viscous forces dominate over shorter timescales and are the ones in charge of ultimately decelerating the object \cite{Bilotto:2024}. 
Quasi-static solid deformation means that the elastic wave propagation time $\tau_{\mathrm{prop}}$ across the indenter tip is much shorter than the timescale of lubrication pressure buildup $\tau_{\mathrm{impact}}$; i.e., using the dimensionless parameter $\phi = \tau_{\mathrm{prop}} / \tau_{\mathrm{impact}}$ introduced by Bilotto et al.~\cite{Bilotto:2024}, $\phi \ll 1$. 
In the limit $\phi \ll 1$, elastic stresses can equilibrate quickly and the solid responds quasi-statically to the evolving fluid pressure. 
The dry contact mechanics results for comparison are taken from the literature \cite{Sneddon} and verified with boundary-element method simulations \cite{Tamaas}.

\begin{figure*}[t]
\centering

\begin{subfigure}[t]{0.22\textwidth}
    \centering
    \caption{\centering \qquad \Large $n = 1$}
    \includegraphics[width=\linewidth]{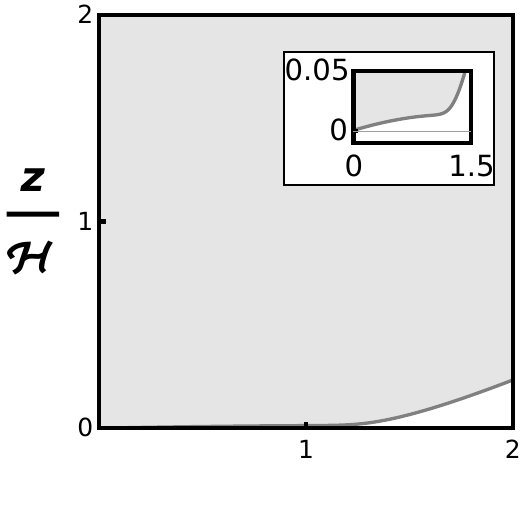}
    \includegraphics[width=\linewidth]{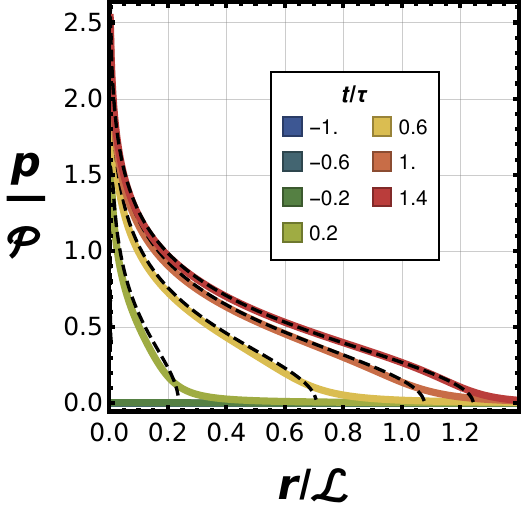}
\end{subfigure}
\hspace{1pt}
\begin{subfigure}[t]{0.22\textwidth}
    \centering
    \caption{\centering  \qquad \Large $n = 2$}
    \includegraphics[width=\linewidth]{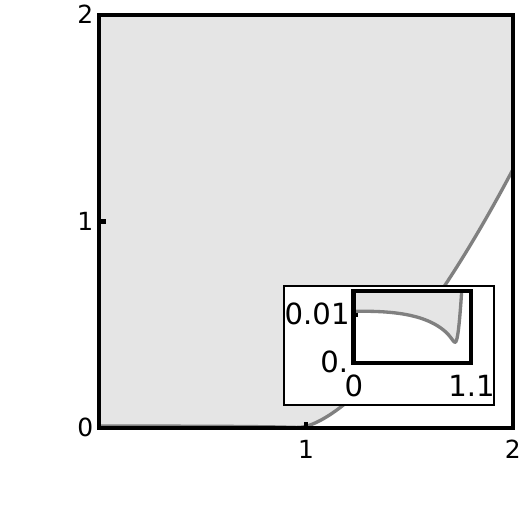}
    \includegraphics[width=\linewidth]{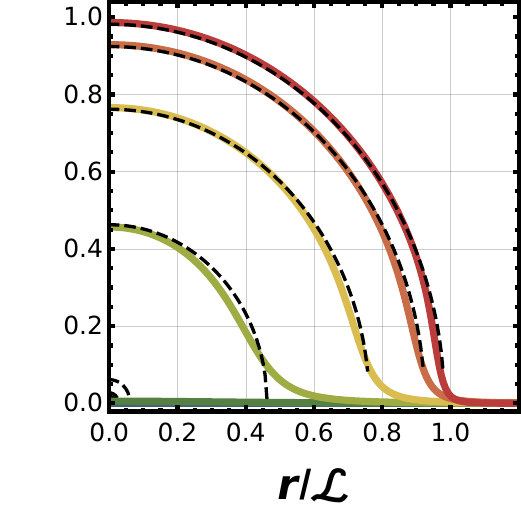}
\end{subfigure}
\hspace{1pt}
\begin{subfigure}[t]{0.22\textwidth}
    \centering
    \caption{\centering \qquad  \Large $n = 3$}
    \includegraphics[width=\linewidth]{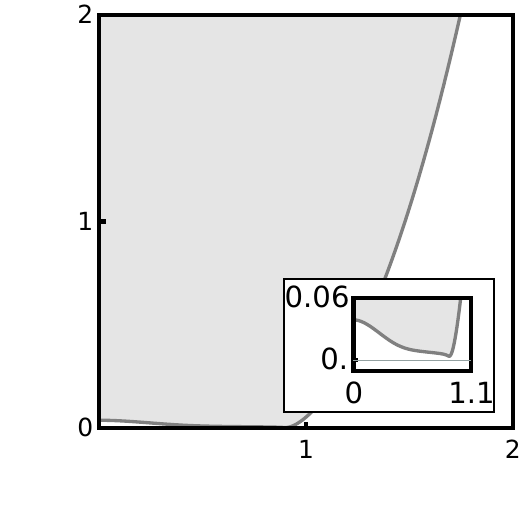}
    \includegraphics[width=\linewidth]{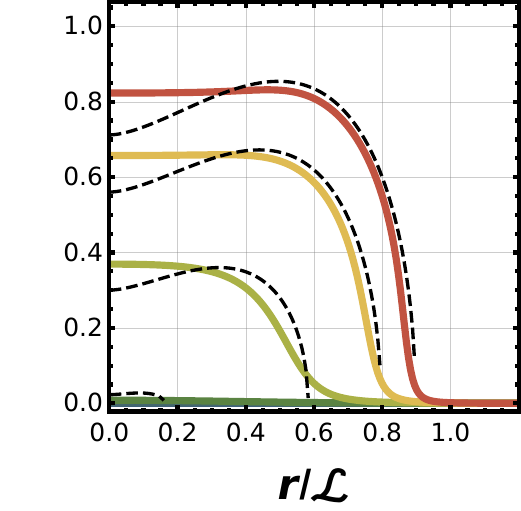}
\end{subfigure}
\hspace{1pt}
\begin{subfigure}[t]{0.22\textwidth}
    \centering
    \caption{\centering \qquad  \Large $n = 6$}
    \includegraphics[width=\linewidth]{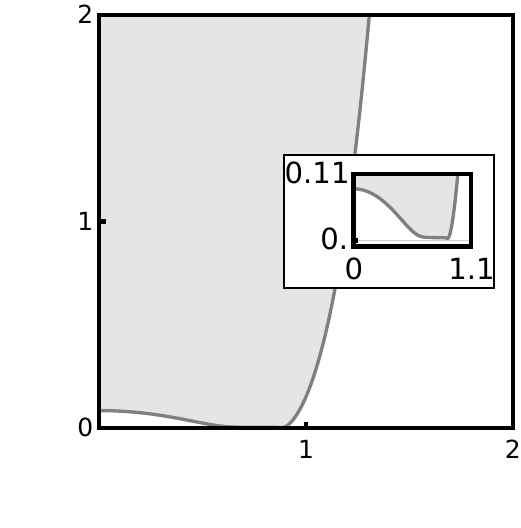}
    \includegraphics[width=\linewidth]{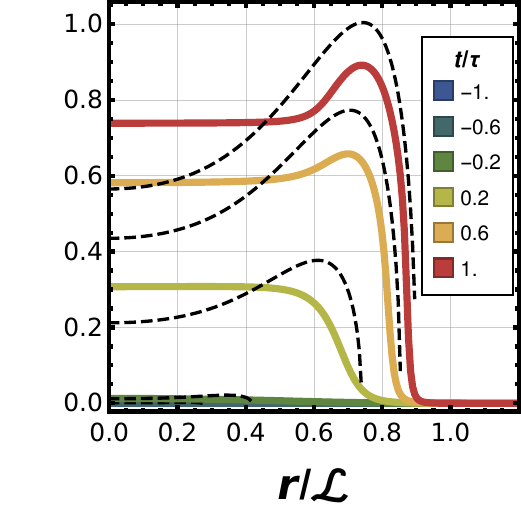}
\end{subfigure}

\caption{
Comparison of fluid-mediated deformation and pressure distribution for different shape exponents $n$. 
Each column shows the final deformed tip profile (dimensionless results) at the top (with a zoom at the tip in the inset), and the corresponding radial pressure distribution at selected times at the bottom. 
Vertical axes correspond to dimensionless height (top row) and dimensionless pressure (bottom row); horizontal axes are always dimensionless radial coordinate. 
Recall that the characteristic values $\mathcal{H}, \mathcal{L}, \mathcal{P}$ depend on $n$ (see Supplementary Material). 
Simulation results are in color and dry contact mechanics (DCM) approximation, based on numerical results of deformation at the tip, are in black dashed lined. 
The latter is computed using the impactor deformation at each time $t$, $\delta_{\text{num}}$, taken to be the value on the axis, i.e., $\delta_{\text{num}} = w(0,t)$. 
For instance: in the parabolic case $f(r) = r^2/2R$, the pressure distribution predicted with DCM is 
    $p(r,t) 
    =
    {2 E^* / \pi}
    (R \delta_{\text{num}}(t) - r^2)^{1/2}$ 
.
Pressure results correspond to maximum pressure (no matter where the peak happens) in all cases but $n=1$, because the pressure is singular there; the finite pressure at $r = 0.04 \mathcal{L}$ is monitored instead. 
The dimensionless time is shifted so that $t/\tau = 0$ would correspond to the moment in which the solid would grace the substrate if there was no mediating fluid. 
The times corresponding to $n=1, 2$ are in the leftmost panel, the ones for $n=3, 4$ are in the rightmost one. 
Sharper tips ($n = 1$) remain convex and make central contact, while blunter shapes deform into dimples that trap air and shift the pressure peak outward. 
}
\label{fig:collage_pressure_bubble}
\end{figure*}

Our results show that the geometry of the indenter tip plays a central role in the formation of contact and the entrapment of air. 
We first examine the deformation of each indenter ($n = 1, 2, 3, 6$) at the end of the approach, when the velocity of the center of mass of the indenter becomes zero and is about to change sign (i.e., the rebound is about to start). 
Thus, Figure~\ref{fig:collage_pressure_bubble} (top row) shows the final deformed profiles at the onset of rebound. 
%
%
%
%
While the conical tip ($n = 1$) remains convex and would contact the surface at its apex, all blunter shapes develop a central dimple. 
The (dimensionless) normalized height and volume of the trapped air pocket increase with $n$. 
%

So, under quasi-static conditions ($\phi \ll 1$) and thin-layer lubrication forces ($\mathrm{St} \gg 1$), tip sharpness controls the likelihood of air entrapment: conical profiles would maximize solid-solid contact and avoid the formation of a dimple at the tip, while rounded shapes promote bubble entrapment. 
The fluid film mediates deformation in a way that is remarkably well predicted by dry contact mechanics, provided the deformation remains convex. 
For sharp profiles ($n \leq 2$), the approach leads to a radially monotonic pressure buildup and full contact at the center, with negligible air entrapment. 
In contrast, for rounded profiles ($n > 2$), the pressure peak shifts away from the axis. 
The divergence between the numerical elastohydrodynamic results and dry contact mechanics is in all cases attributed to the formation of the dimple: 
initially, the profile flattens, so it is undergoing a very similar deformation (and hence the same interface tractions) as \Cref{fig:schematic}(b), but by the end of the process their deformations diverge, meaning that after flattening the profile, the lubrication pressures still form a dimple on the solid surface, thus the pressure distributions must differ. 
The relatively larger the bubble, the larger the departure from the dry contact mechanics solution. 


To compare the actual amount of fluid entrapped for different indenter shapes, we ``normalize'' the systems by imposing matching tip geometry at a fixed point (\Cref{fig:indenters}) and identical material properties across all profiles: mass $m$ and contact modulus $E^* = E /(1 - \nu^2)$, $E$ being the Young's modulus of the material and $\nu$ is its Poisson ratio. 
Specifically, we ensure that the parabolic, cubic, and hexic indenters intersect at the same radial location and height, $z = R/10$ for $r = R/\sqrt{5}$. 
%
%
This condition provides consistent values of the geometric constant $k$ for each shape and allows direct comparison of the bubble volumes.

The volume of entrapped air, $\Omega_b$, can be expressed in terms of the characteristic vertical and radial length scales as $\Omega_b = \tilde{\Omega}_b \mathcal{H} \mathcal{L}^2$, where the dimensionless bubble volumes $\tilde{\Omega}_b$ are obtained from the numerical simulations. 
Let us compare the cubic and the hexic to the parabolic profile by computing 
$\Omega_b^{(n)} / \Omega_b^{(2)}$ for $n=3$ and $n=6$. 
These ratios can be shown to depend on the dimensionless group $\Pi = mV^2 / E^* R^3$; 
%
it follows that
$\Pi \sim (V/c)^2$ where $c$ is either wave speed in the bulk material. 
For a detailed derivation, we refer the reader to the Supplementary Material. 

Table~\ref{tab:bubble_volume_comparison} shows the predicted relative bubble volumes for two representative values of $c/V$, corresponding to a soft and a stiff material. We find that for both cases, blunter profiles lead to significantly larger trapped volumes: up to three times more for $n=3$, and over ten times more for $n=6$ in stiffer materials. 
This reveals the consistent trend: sharper shapes yield smaller cavities and more contact around the tip.

\begin{table}[H]
\centering
\caption{Relative bubble volume $\Omega_b^{(n)} / \Omega_b^{(2)}$ for cubic ($n=3$) and hexic ($n=6$) indenters, compared to the parabolic case, for two material stiffness levels.}
\label{tab:bubble_volume_comparison}
\begin{tabular}{c|c|c}
$n$ & $c/V = 100$ (soft) & $c/V = 1000$ (stiff) \\
\hline
3 & 2.20 & 3.27 \\
6 & 5.48 & 12.81 \\
\end{tabular}
\end{table}

If one revisits the results for a locally-parabolic indenter ($n=2$) in Ref.~\cite{Bilotto:2024}, a striking resemblance stands out between the dimple shapes in the insets of \Cref{fig:collage_pressure_bubble} for $n=3, 6$ and the dimple shape forming in the former case when the quasi-static assumptions breakdowns, $\phi \sim 1$. 
This similarity between the effects of tip shape and approach velocity hints a deeper connection between the two parameters in controlling fluid-mediated deformation. 
A blunter indenter moving slowly may generate the same pressure and deformation fields as a sharper one moving faster. 
To examine this idea, let us define a ``lubrication pressure front'' as the radial extent over which the interface pressure $\sim \mathcal{P}$, and consider how it propagates radially during approach and how that propagation couples to the geometry of the leading edge.
The pressure front develops as regions of the edge ``fall into'' the thin film and begin to experience the viscous resistance. 
Assuming a point enters the pressure front when its gap height reaches a threshold $h^*$ (ignoring deformation), the radial location of the front $r^*(t)$ evolves according to 
\[
h^* = h_0 - V t + \frac{r^n}{2k} \quad \Rightarrow \quad r^*(t) = \left[ 2k (h^* - h_0 + Vt) \right]^{1/n}.
\]
Differentiating, the radial velocity of the front becomes
\[
\dot{r}^*(t) = \frac{(2kV)^{1/n}}{n} t^{(1 - n)/n}.
\]
For $n > 1$, this velocity diverges as $t \to 0$, implying that the pressure front sweeps radially outward extremely rapidly at early times, much faster than information can be dispersed through the solid by elastic waves. 
Thus, in effect, the fluid locally sees a ``flat punch'' \cite{Popov:2019}, regardless of the nominal shape. Only after a finite delay, $\sim \Delta T^*$, the front does slow down enough for the leading edge of the solid to start to resist the pressure in coherent fashion. 
To estimate this delay, we define $\Delta T^*$ as the time at which the front speed matches the wave speed $c$, giving
\[
\dot{r}^*(\Delta T^*) = c \quad \Rightarrow \quad \Delta T^* \approx \left( \frac{(2kV)^{1/n}}{n c} \right)^{n/(n - 1)}.
\]
Comparing this time span from shape to shape is particularly revealing (showing $n=3$ and $n=2$ for illustration):  
\[
    {\Delta T^*|_{n=3} \over \Delta T^*|_{n=2}}
    \sim
    \sqrt{c \over V} \gg 1 \, ,
\]
so, as intuitively one would expect, the time that it takes to ``for the pressure to couple with shape'', or to ``see local curvature'', is much shorter in the parabolic case than in the cubic one, because the latter is flatter around the tip so most of its leading edge engages with the lubrication pressure around the same instant. 
To assess how significant this effect is relative to the overall dynamics, we compare $\Delta T^*$ to the global approach timescale $\tau \sim \mathcal{H}/V$. 
Using the dry mechanics scales again, this yields:
\[
\frac{\Delta T^*}{\tau} \sim \left( \frac{V}{c} \right)^{(n+1)/(n+1/2)} \ll 1.
\]
This ratio remains small for all practical values of $n$ and $V$, but its direct proportionality with $V$ and $n$ is coherent with the trend we expect: blunter shapes and higher velocities lead to most of the leading edge engaging around the same time, much quicker than surface waves can propagate over the profile. 
Conversely, sharper shapes or slower approach velocities allow the pressure front to couple to the geometry.

Interestingly, the conical shape ($n = 1$, in this case $2k=\cot \alpha $) stands out in this context. 
Its pressure front speed is constant in time, $\dot{r}^* = V \cot \alpha$, independent of $t$, and the transition between “seeing” or “not seeing” the tip curvature depends only on the ratio $V/c$ and the cone angle $\alpha$. 
In the quasi-static limit, this shape does not lose its convexity during approach, recall \Cref{fig:collage_pressure_bubble}. 
However, as $V$ increases, the pressure front may outpace elastic signals for sufficiently shallow cones (small $\alpha$), potentially leading to dimple formation even in the conical case. 
This suggests that conical tips are optimally robust against air entrapment under slow conditions, but could still exhibit effective “blunting” at high velocity, depending solely on material stiffness and cone angle. 

To fully confirm the picture outlined in previous paragraphs, either experiments or computer simulations beyond the quasi-static regime ($\phi \to 1$) need to be conducted.

Under quasi-static conditions ($\phi \ll 1$) and localized thin-film viscosity ($\mathrm{St} \gg 1$), we have shown that the shape of a deformable object significantly influences the onset of contact and the entrapment of fluid during its approach to a rigid substrate through a viscous film. 
For axisymmetric tips described by power-law profiles $f(r) \sim r^n$, our simulations reveal that sharper shapes (lower $n$) maximize contact area and reduce or eliminate air entrapment, while blunter shapes promote the formation of central dimples and trapped bubbles. 
Remarkably, the evolving pressure distribution in the fluid-mediated system matches well that predicted by dry contact mechanics, but the match deteriorates the blunter the profile. 
This is attributed to ever larger dimples forming when the leading edge is blunter, which renders the deformation significantly different from the simple flattening predicted by dry contact mechanics.
Among the profiles considered, the conical tip ($n = 1$) is the only one that remains convex throughout the approach and contacts at its apex, suggesting it as an optimal geometry for full contact in slow, fluid-mediated interactions (elastohydrodynamic lubrication). However, we have argued that even this geometry may entrap air if the quasi-static solid-response assumption loses its validity. 

These findings motivate several future work directions. 
Simulations in the dynamic regime ($\phi \sim 1$), where elastic wave propagation and lubrication evolve on similar timescales~\cite{Bilotto:2024}, would test the limits of the shape–velocity analogy and probe inertial effects. 
A deeper analysis of bubble geometry, as in Ref.~\cite{bertin:2024}, or the behavior during rebound and adhesion~\cite{bertin:2025} may further reveal the influence of shape beyond just elastohydrodynamic approach. 
Finally, experimental exploration of non-parabolic tip geometries, building on studies such as Ref.~\cite{Zheng:2021}, would provide a valuable test of the theoretical framework developed here.


\textbf{Acknowledgements}.  
I am thankful to Prof. J.-F. Molinari and Mr. J. Bilotto (EPFL) 
for stimulating discussions. 
The support of the Swiss National Science Foundation via Ambizione Grant 216341 ``Data-Driven Computational Friction'' is gratefully acknowledged.

\textbf{Data availability}. 
All data will be made available openly before publication. 
The supplementary material file can be retrieved from the repository \texttt{github.com/jgarciasuarez/soft-contact}.

\bibliography{references}

\end{document}